# Molecular Water Lilies: Orienting Single Molecules in a Polymer Film by Solvent Vapor Annealing

*Dominik Würsch[1], Felix J. Hofmann[1], Theresa Eder[1], A. Vikas Aggarwal[2], Alissa Idelson[2], Sigurd Höger[2], John M. Lupton[1], Jan Vogelsang[\*,1]*

[1]Institut für Experimentelle und Angewandte Physik, Universität Regensburg, 93053 Regensburg, Germany

[2]Kekulé-Institut für Organische Chemie und Biochemie, Universität Bonn, 53121 Bonn, Germany

ABSTRACT:

The microscopic orientation and position of photoactive molecules is crucial to the operation of optoelectronic devices such as OLEDs and solar cells. Here, we introduce a shape-persistent macrocyclic molecule as an excellent fluorescent probe to simply measure (i) its orientation by rotating the excitation polarization and recording the strength of modulation in photoluminescence (PL), and (ii) its position in a film by analyzing the overall PL brightness at the molecular level. The unique shape, the absorption and the fluorescence properties of this probe yields information on molecular orientation and position. We control orientation and positioning of the probe in a polymer film by solvent vapor annealing (SVA). During the SVA process the molecules accumulate at the polymer/air interface, where they adopt a flat conformation, much like water lilies on the surface of a pond. The results are significant for OLED fabrication and single-molecule spectroscopy (SMS) in general.

TOC Figure:

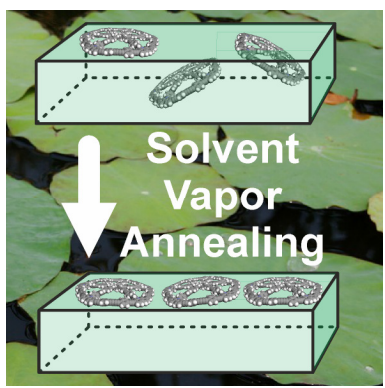

The brightness of an organic light emitting diode (OLED) depends on various processes and parameters.[1-3] The most important ones are: charge injection at the interfaces,[4] charge mobility,[5] photoluminescence (PL) quantum yield,[6] and light out-coupling efficiency.[7] Single-molecule spectroscopy has already been applied successfully to a few of these aspects to find molecular structure-property relationships,[8-10] but has been mainly neglected in the context of light out-coupling efficiency in OLEDs. This very important issue is influenced by the orientation of the transition dipole moment (TDM) of the luminescent molecules with respect to the film plane, because of waveguiding effects in the emissive layer.[11-14] It was found that a molecular orientation horizontal with respect to the substrate surface is advantageous, because increased charge carrier mobility can be reached and coupling into lossy waveguide modes is minimized.[15] Nowadays, the luminescent molecules are embedded within a host matrix, which is mainly responsible for the charge transport.[16] The concentration of the luminescent molecules is chosen to be low (1-10 %) to avoid aggregation-induced self-quenching effects. Such effects can be present in conjugated polymers, where they arise due to excitonic coupling[17] and efficient intermolecular energy transfer.[18] It is therefore desirable to measure and even control the orientation of isolated luminescent molecules embedded in a host matrix. Fortunately, in single-molecule spectroscopy (SMS), small luminescent organic molecules are typically observed by embedding them into non-fluorescent thin polymer host matrices.[19-20] Such an approach yields similar conditions to those encountered in OLEDs and other thin-film organic optoelectronic devices.

Several publications dealing with the PL of immobilized single molecules embedded in a thin polymer film demonstrate the impact of the TDM orientation on various observables, e.g., the PL lifetime,[21] PL intensity[22] or the degree of spatial localization of the PL emission.[23-24] The latter is of utmost importance for super-resolution microscopy: it was shown by Moerner *et al.* that a random orientation of the TDM can lead to severe deterioration in localization-based super-resolution microscopy methods.[23] For this reason, elaborate microscopy techniques have evolved to determine the exact 3D-orientation and position of the single-molecule TDM,[25] based on defocused imaging[26] and applying an elaborate correction to the point-spread function,[27] respectively.

Here, we demonstrate that a certain molecule with particularly beneficial geometry and photophysical properties can be used without elaborate microscopy techniques to determine its orientation and position at the molecular level. The symmetry of a spoked wheel-shaped molecule is exploited in combination with excitation polarization fluorescence spectroscopy. Its triplet-state photophysics is utilized as a reporter on the position. Further, we demonstrate that the orientation as well as the position of such molecules can be homogenized by subsequently applying solvent vapor annealing (SVA), an industrially important post-processing technique for the equilibration of thin polymer films.[28-29] We show that the molecules orient themselves parallel to the surface and accumulate at the polymer/air interface. On the one hand the findings are potentially of importance for the design of OLEDs, on the other hand they play a major role for the field of SMS where the molecules under investigation are embedded in a thin polymer film. We suggest that instead of applying sophisticated microscopy techniques to determine the 3D-orientation and position of single-molecule TDMs to correct for their impact on spectroscopic characteristics, it is possible to homogenize the TDMs of all molecules by SVA to directly compare their spectroscopic features in standard fluorescence microscopes.

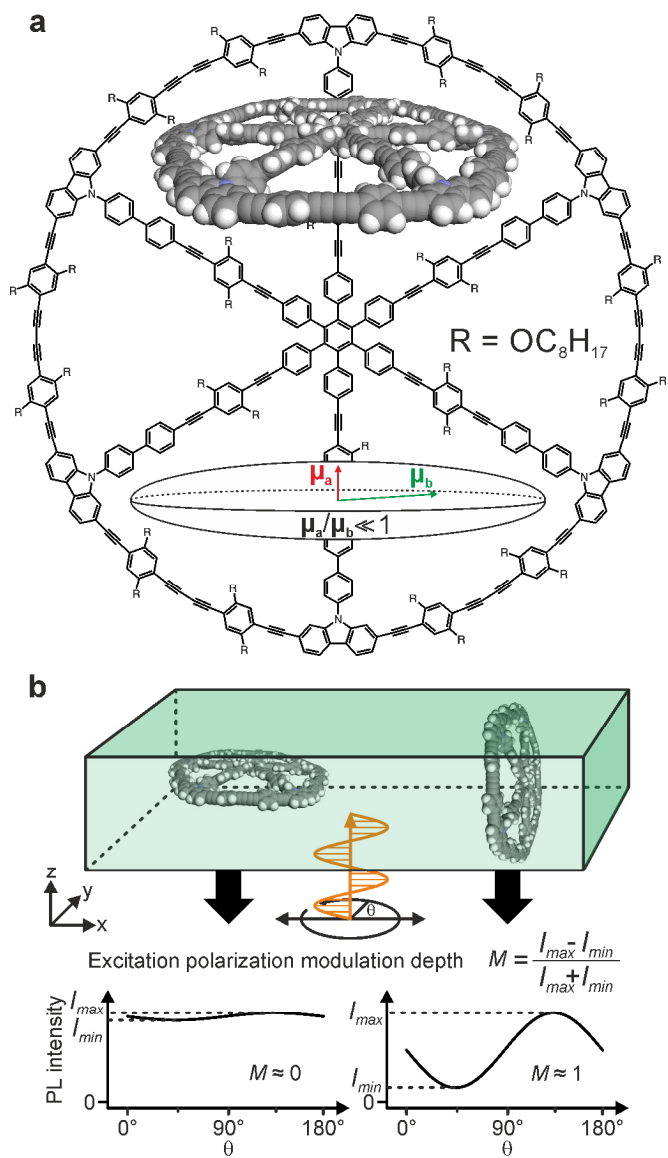

**Figure 1.** Chemical structure of the fluorescent molecule and the measurement procedure to determine single-molecule orientation. (a) The chemical structure of the shape-persistent spoked-wheel molecule is drawn including a volumetric model in the top and a schematic of the 3D-absorption ellipsoid in the bottom, with $\mu_a$ and $\mu_b$ corresponding to transition dipole moment (TDM) vectors. (b) Schematic drawing of the excitation polarization fluorescence spectroscopy measurements and definition of modulation depth, $M$. The highly anisotropic form of the absorption ellipsoid yields low $M$-values for molecules oriented parallel to the xy-plane, and high

*M*-values for orthogonal orientations. The *M*-value is therefore a measure of the orientation and not of molecular shape, as is usually the case.

The molecular orientation is extracted as follows. In a nutshell, we perform a 3D-projection of the absorption ellipsoid of a shape-persistent wheel molecule onto a 2D-plane. If the molecule is oriented flat in the 2D-plane, a isotropic absorption with respect to the excitation polarization will be measured, whereas anisotropic absorption is expected for molecules with perpendicular orientation. The chemical structure of the molecule is shown in Figure 1a, with a volumetric model inset in the top and the respective absorption ellipsoid shown at the bottom. The diameter of the molecule was previously measured by STM to be ~6 nm.[30] The shape of the absorption ellipsoid is determined by the ratio $r = |\mu_a|/|\mu_b|$, where $\mu_a$ and $\mu_b$ are TDM vectors.[31] The photophysics and synthesis of this material were reported by Aggarwal *et al.* and Thiessen *et al.*, who found that the rim is the optically active region and not the spokes.[30, 32] In excitation, the wheel symmetry is preserved because incident light of any polarization can be absorbed in the case of molecules oriented in the plane. In emission, the exciton localizes randomly onto different segments, resulting in overall unpolarized PL.[30] Hence, for this molecule the ratio $r$ for the absorption ellipsoid is close to zero. The projection of the absorption ellipsoid onto a 2D-plane is measured by excitation polarization fluorescence spectroscopy as shown in Figure 1b.[33]

The molecules are embedded in a thin film of poly(methyl-methacrylate) (PMMA) at single-molecule concentrations, by dynamically spin coating a 5 % w/w PMMA/toluene solution with a molecule concentration of ~$10^{-12}$ M at 2,000 rpm for 120 s. This procedure yields a film of ~250 nm thickness with an average density of ~40 individual molecules over an area of 50×50 µm². A schematic drawing of the film is shown in the top part of Figure 1b including two molecules with different orientations with respect to the interface. The linearly polarized excitation beam is focused onto the back focal plane of the objective, which results in a large excitation area of ~80×80μm² and negligible polarization in the z-direction orthogonal to the imaging plane. The excitation area is imaged onto an EMCCD camera and the PL of single molecules is recorded while the excitation polarization is rotated in the xy-plane of the sample. The excitation polarization modulation depth, $M$, is obtained by fitting the PL intensity, $I$, as a function of polarization angle, $\theta$, to $I(\theta) \propto 1 + M\cos 2(\theta - \Phi)$, where $\Phi$ is the orientation angle for maximal PL intensity. $M$ can be also described by the difference of maximum and minimum PL divided by the sum, as stated in Figure 1b. The 3D-projection of the wheel molecules onto the xy-plane yields $M$-values close to zero for the case of molecules lying flat (bottom left of Figure 1b) and $M$-values close to one for molecules of perpendicular orientation (bottom right of Figure 1b). Because absorption and emission of the wheel structure are fundamentally unpolarized, the $M$-value is a direct measure of the orientation of this molecule. Note that $M$ is usually invoked as a metric of molecular morphology. This approach can, however, be quite misleading if the orientation of the molecule is not known. For our molecules this ambiguity is avoided, since all molecules have the same shape due to their extraordinary rigidity.[34] $M$ of the molecules used here therefore reports solely on molecular orientation.

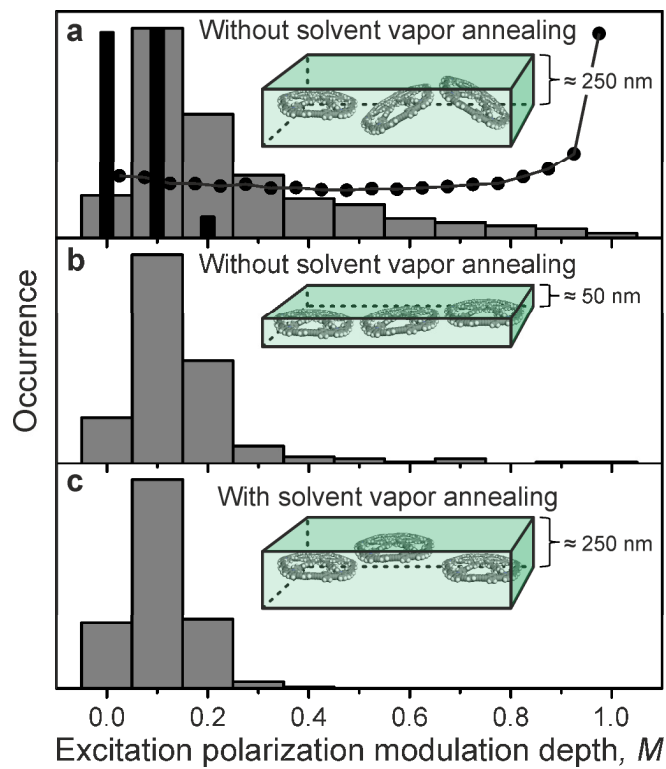

**Figure 2.** Excitation polarization modulation depth, $M$, histograms of molecules embedded in poly(methyl-methacrylate) (PMMA) films. (a) $M$-histogram of 15 measured areas (50×50 µm²) with a total of 750 molecules without any post processing. The connected black dots correspond to the expected distribution of $M$-values for arbitrary orientations of molecules in three dimensions as obtained by a Monte-Carlo simulation. The black bars correspond to the instrument response for excitation of an isotropic fluorescent bead with similar PL intensity as the molecules. (b) $M$-histograms of 6 measured areas with a total of 227 molecules embedded in a 50 nm PMMA film and (c) of 10 measured areas with a total of 425 molecules in a 250 nm film with subsequent annealing in a toluene-saturated nitrogen gas. The insets are a schematic representation of the PMMA films with the corresponding orientations of the molecules.

The $M$-values of several hundred single molecules are measured by accumulating the $M$-values of multiple measured areas (50×50 µm²) and the corresponding histograms for different sample preparation conditions are shown in Figure 2. The spot density of ~40 molecules per 50×50 µm² is similar for all samples. A theoretically expected distribution of *M*-values is plotted as black dots in Figure 2a under the assumption that the orientation of the molecules is completely arbitrary. This distribution of randomly oriented molecules was obtained by a Monte-Carlo simulation of 10,000 absorption ellipsoids with $r = 0$ following the procedure described previously.[35] More values for $M \approx 1$ are generated than for $M \approx 0$ as one would expect intuitively: $M \approx 0$ can only be obtained for absorption ellipsoids which lie flat. These have one degree of freedom, rotation around the central axis of the absorption ellipsoid which is parallel to $\mu_a$. $M \approx 1$ is obtained for absorption ellipsoids oriented perpendicular to the imaging plane, which have an additional degree of freedom: rotation around an axis perpendicular to the central axis of the absorption ellipsoid, i.e. parallel to $\mu_b$ and perpendicular to the xy-plane.

The grey bars in Figure 2a display the *M*-histogram of molecules embedded in a ~250 nm thick PMMA film (spin-coated with a 5 % w/w PMMA/toluene solution at 2,000 rpm) without any subsequent post-processing. To test the accuracy of the measurement, we superimpose a histogram (black bars) of fluorescent polystyrene beads of similar spectral properties and PL intensities. Most of the molecules are oriented flat in the sample plane because mainly low *M*-values are measured around 0.1, but a long tail reaching towards higher *M*-values up to 1 is also present, indicating a non-negligible sub-population of molecules perpendicular to the plane. A similar distribution of orientations has previously been observed for conjugated polymers by defocused widefield imaging.[36] However, the reason for the predominantly flat orientation on the substrate surface is still unclear and it was previously speculated that this observation arises mainly due to shear forces during spin-coating.[37] On the one hand, this hypothesis is not supported by our data because no differences for the *M*-value distribution is observed for different regions of the film, which should be subject to varying strengths of shear forces. On the other hand, larger shear forces may be expected for a ~50 nm PMMA film than for the thicker film, and our results appear to support this hypothesis. The frequency of larger $M$-values ($M \geq 0.3$) of the molecules in Figure 2a is significantly reduced by embedding the molecules in a ~50 nm film (by spin-coating a 1 % w/w PMMA/toluene solution at 2,000 rpm), as the $M$-histogram in Figure 2b demonstrates. Here, almost all molecules are now oriented flat in the sample plane. However, Figure 2c shows that a very similar result can be achieved even for the ~250 nm film treated by SVA. A home-built gas-flow chamber was used to employ a toluene saturated nitrogen atmosphere to swell the PMMA film (details of the SVA procedure can be found in refs.[38-39]). During SVA, the PMMA film is in a heterogeneous mixture of solid- and liquid-like phases in which the single molecules can diffuse.[38] This process is stopped after 5 minutes and the film dried again in a pure nitrogen atmosphere. According to Hoang *et al.* the PMMA film is not completely swollen during the first 10 minutes, which were employed here.[40] However, diffusion of the molecules sets in already a few minutes after the SVA process is started and we stop the SVA process before the PMMA film is completely swollen since aggregation should be avoided. On average, the same number of molecules were counted per 50×50 µm² area pre and post processing, suggesting that aggregation during SVA can be neglected. The average orientation of the molecules is heavily impacted by SVA because no $M$-values above 0.5 are measured (see Figure 2c) in the ~250 nm film after SVA. Since no shear forces are present during SVA, we conclude that these forces cannot be responsible for the predominantly flat orientation of the molecules, in contrast to previous claims.[37] A direct comparison with the $M$-histogram for completely unpolarized beads (black bars, Figure 2a) reveals on average still slightly higher $M$-values of the ring molecules which can be an indication for a "boat conformation", as was reported previously for a similar ring structure.[34] Next, we investigate the position of the molecules in the thin film and the impact of SVA, which also gives a strong indication of the molecular mechanism behind the re-orientation.

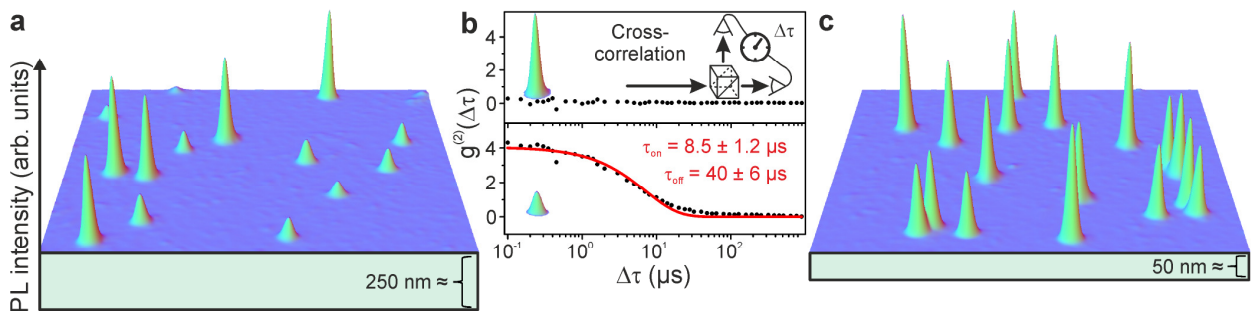

**Figure 3.** Confocal scanning microscopy images and second-order cross-correlation, $g^{(2)}(\Delta\tau)$, analysis. (a) 10×10 µm² confocal scan image of molecules embedded in a 250 nm thick PMMA film without subsequent SVA. Two populations of single molecules are seen as bright and dim spots. (b) Accumulation of $g^{(2)}(\Delta\tau)$ curves for 39 diffraction limited spots with high PL intensity (top panel) and 254 spots with low PL intensity (bottom panel). The $g^{(2)}(\Delta\tau)$ curve for the low-intensity PL molecules is fitted by a single exponential decay (red curve) from which the average on-time, $\tau_{on}$, and off-time, $\tau_{off}$, are extracted. (c) 10×10 µm² confocal scan image of molecules embedded in a 50 nm thin PMMA film. All single molecules have the same brightness.

As for most fluorescent organic molecules, the PL of the molecules studied here is also extremely sensitive to the oxygen concentration of the surrounding environment (see Figure S1).[41] It is well-known that oxygen effectively quenches the triplet state.[42] In an oxygen-free environment the triplet state becomes long-lived with a lifetime ranging from µs up to even seconds.[43-44] As a consequence, the PL intensity can be saturated under increased excitation densities, owing to shelving in the long-lived triplet state.[18] We hypothesize that molecules deeply embedded within the PMMA film have less access to molecular oxygen as compared to fluorophores close to or even at the PMMA/air interface. Consequently, molecules inside the film should have a longer triplet-state lifetime and exhibit less overall PL intensity than molecules at the surface. To test this hypothesis we inspected the photophysics of molecules inside PMMA films of ~250 nm and ~50 nm thickness, respectively, by confocal fluorescence microscopy under ambient conditions. First, the higher excitation intensity in confocal microscopy of ~0.5 kW/cm², as compared to the wide-field illumination described above at ~1 W/cm², leads to more frequent shelving of the molecule in the triplet state. Second, avalanche photodiodes (APDs) are used in a Hanbury Brown-Twiss geometry to measure the second-order cross-correlation function, $g^{(2)}(\Delta\tau)$, of the PL signal of single molecules with high time resolution as sketched in the top panel of Figure 3b. Confocal scan images of 10×10 µm² size were measured for molecules embedded in a ~250 nm PMMA film, and diffraction-limited spots are observed with distinct different brightness. One such image is shown in Figure 3a as a 3D-contour plot to visualize the different PL intensities. A histogram of the PL intensities of 764 spots reveals a bimodal distribution of dim and bright spots (Figure 4a). The photophysics of the bright and dim spots was further inspected by placing the molecules in the confocal excitation spot and subsequently recording the PL until the molecule undergoes irreversible photobleaching. Next, $g^{(2)}(\Delta\tau)$ was measured for each molecule and plotted for an accumulation of 39 bright spots and 254 dim spots in the top and bottom panel of Figure 3b, respectively. Whereas the bright spots show no discernible PL intensity fluctuations from 0.1 µs up to 1 ms as demonstrated by a flat $g^{(2)}(\Delta\tau)$ curve, the dim spots exhibit a pronounced amplitude of $g^{(2)}(\Delta\tau)$, which corresponds to strong PL intensity fluctuations.[45-46] In the case of the dim spots, $g^{(2)}(\Delta\tau)$ can be well fitted with a single-exponential function of the form $g^{(2)}(\Delta\tau) = A \cdot \exp(-\Delta\tau/\tau_0)$, given as the red curve in Figure 3b.

The amplitude $A = 4.1 \pm 0.2$ and the characteristic decay time $\tau_0 = 7 \pm 1$ µs are extracted. The single-exponential decay of the cross-correlation suggests that only one dark state, namely the triplet state, is involved in the PL intensity fluctuations with time.[18, 45-46] Before extracting the average off-time of the fluorescence, $\tau_{off}$, which corresponds to the lifetime of the triplet state, and the average on-time, $\tau_{on}$, which describes the time the molecule resides in the singlet manifold, the amplitude $A$ must be corrected for the measured signal to background ratio, $S/B \approx 15$, of the PL intensity according to $A_{corr} = \left(\frac{S+B}{S}\right)^2 \cdot A$, which yields $A_{corr} = 4.7 \pm 0.2$.[46] With the obtained values of $A_{corr}$ and $\tau_0$, $\tau_{on}$ and $\tau_{off}$ are calculated from $\tau_{on} = \tau_0 \cdot (1 + 1/A_{corr})$ and $\tau_{off} = \tau_0 \cdot (1 + A_{corr})$, yielding $\tau_{on} = 8.5 \pm 1.2$ μs and $\tau_{off} = 40 \pm 6$ μs, respectively.[46] The dim spots only reside in the singlet manifold for ~18 % of the measured time and are in the triplet state for the remaining time (82 %). In contrast, the bright spots spend effectively ~100 % of the observation time in the singlet manifold. For this reason, we expect the dim spots to exhibit, on average, a fivefold reduction in PL intensity as compared to the bright spots. This conclusion from the temporal dynamics of photon emission is in perfect agreement with the measured peak positions of the PL intensity distributions of the dim and bright spots, shown in Figure 4a. The PL intensity distribution of the spots is bimodal, with the ratio between the two distributions being 5 and the number of bright (dim) molecules is counted to be 486 (278). Most importantly, the molecules embedded in the 50 nm film exhibit almost exclusively high PL intensities, which is reflected by the number of bright (dim) molecules of 871 (110) and no fluctuations in PL intensity in the cross-correlation experiment. The PL image in Figure 3c shows that all spots have approximately the same brightness, which is plotted in more detail in the PL intensity histogram in Figure 4b. It is important to note that no differences in the PL lifetime and the PL spectrum are observed between the dim and bright spot populations (see Figure S2), indicating that there is no significant impact of the surface on the excited-state lifetime and, more importantly, that the molecules are all intact. We conclude that the PL intensity is a direct observable to distinguish between molecules embedded deep within a film, where access to molecular oxygen is limited, and molecules at the PMMA/air interface with sufficient access to oxygen to quench the triplet state. The bimodal distribution also suggests that the oxygen concentration must change very abruptly at the surface. Additionally, due to the perfect agreement between the dynamics of triplet state shelving and the intensity ratio between bright and dim spots, we can safely assume that surface effects, such as changes of the dipole radiation pattern[22] and TDM orientation effects at the interface,[21] can be neglected here from having an impact on the overall PL intensity.

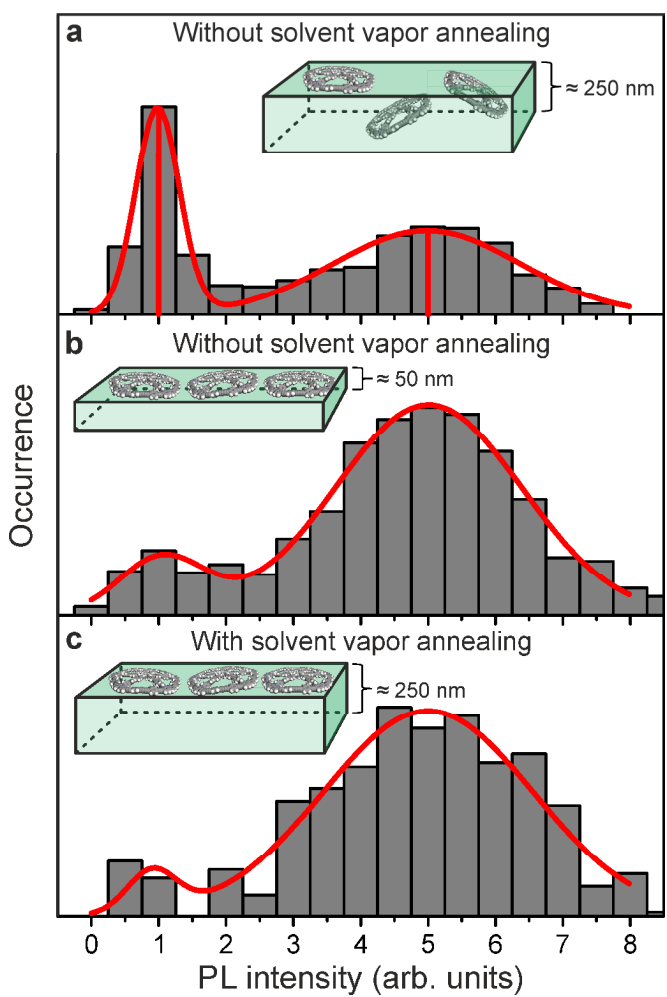

**Figure 4.** PL intensity histograms of molecules embedded in PMMA obtained by confocal scanning microscopy and subsequent analysis of the diffraction-limited point spread functions. (a)

Molecules embedded in a 250 nm thick PMMA film without subsequent SVA yield a bimodal PL intensity distribution, which can be described by a double Gaussian function (red curve). The average intensity ratio of the bright-to-dim population is 5 and the number of bright (dim) molecules is 486 (278). (b) and (c) display PL intensity histograms of molecules embedded in a 50 nm PMMA film and a 250 nm PMMA film with subsequent SVA with the number of bright (dim) molecules of 871 (110) and 385 (18), respectively. The average number of diffraction limited spots per area is similar for all three samples. 764, 981 and 403 molecules were measured for (a), (b) and (c). The insets are corresponding schematics of the molecules distributed in the PMMA films.

Finally, we studied the impact of SVA on the molecular position by measuring the PL intensity. Figure 4c shows the PL intensity histogram of molecules embedded in a 250 nm PMMA film with subsequent SVA carried out for a duration of 5 minutes. The average number of diffraction-limited spots per area remains the same compared to the sample without subsequent SVA, but the population of dim spots is transferred to bright spots as is observed by the number of bright (dim) molecules of 385 (18). SVA of a thin PMMA film with toluene-saturated nitrogen gas is known to lead to strong diffusion of embedded conjugated polymers.[38] The results imply that the molecules diffuse due to Brownian motion randomly through the PMMA film until they reach the PMMA/air interface by chance, where they stay because the swollen PMMA matrix constitutes a less favorable environment as compared to the PMMA/air interface for the embedded molecules. Similar results are also obtained for an almost linear hexamer consisting of the same repeat units as the rim of the wheel molecules (see Figure S3). We conclude that the observations discussed here are not exclusively limited to wheel-shaped molecules. However, in more flexible non-templated linear structures it is harder, if not impossible, to disentangle microscopic orientation and molecular shape in polarization anisotropy data. Further, by combining the observations of Figure 2c and 4c it can be concluded that the molecules minimize their surface interaction with the swollen PMMA environment: they lie down flat on the surface, much like water lilies on the surface of a pond. Aggregation of the mobile molecules can be avoided by applying SVA for only a short time period, e.g. 5 minutes here, because the molecules only need to diffuse through the ~250 nm thick PMMA film in the lateral direction. Longer SVA times can lead to significant aggregation, as previously demonstrated for conjugated polymers.[39] A series of PL intensity histograms of samples subjected to different durations of SVA is shown in Figure S4 and demonstrates that after 5 minutes almost all dim spots have vanished from the distribution. This last point is of substantial importance for improving OLED efficiencies by SVA-induced planarization of molecular orientation because aggregation of the fluorescent molecules might cancel the positive effect gained by an increased light out-coupling efficiency of horizontally oriented TDMs.

In conclusion, we have shown that SVA of a thin polymer film with a "good" solvent which swells the host matrix sufficiently to induce diffusion of the guest reporter molecules leads to an accumulation of all embedded fluorescent molecules at the polymer/air interface. Only a short time period of 5 minutes is required for SVA for this process to complete. The average PL intensity of single molecules can be used to distinguish between molecules embedded inside the film and molecules at the polymer/air interface. A second-order cross-correlation analysis of the PL signal reveals that the triplet-state lifetime is only measurable for molecules inside the film, due to their inhibited access to oxygen, and not for molecules on the surface. Accumulation of molecules at the polymer/air interface leads to a planar orientation of the molecules with respect to the surface, which is revealed in experiment through the excitation polarization fluorescence anisotropy of these unique shape-persistent molecules. We suggest that a brief period of SVA of thin films used in OLEDs, in which anisotropic fluorescent emitters are embedded, could lead to an accumulation of the molecules at the interface along with a beneficial planarization in TDM orientation without significant aggregation. Such nanoscale positioning and orientation of molecules could also be useful for single-molecule OLEDs[47] and single-molecule electromodulation devices[48] to reveal the fundamental interplay between charge injection and light generation.

ASSOCIATED CONTENT

**Supporting Information**. Figures S1-S4 are shown in the Supporting Information. This material is available free of charge via the Internet at http://pubs.acs.org.

AUTHOR INFORMATION

**Corresponding Author**

*jan.vogelsang@physik.uni-regensburg.de

**Author Contributions**

The manuscript was written through contributions of all authors. All authors have given approval to the final version of the manuscript.

ACKNOWLEDGMENT

The authors are indebted to the European Research Council for funding through the Starting Grant MolMesON (305020), to the Volkswagen Foundation for continued support of the collaboration, and thank the German Science Foundation for support through the Graduiertenkolleg (GRK) 1570.

The authors thank Dr. Sebastian Bange for providing the Monte-Carlo simulation used in Figure 2a.

# Supporting Information for:

# Molecular Water Lilies: Orienting Single Molecules in a Polymer Film by Solvent Vapor Annealing

*Dominik Würsch[1], Felix J. Hofmann[1], Theresa Eder[1], A. Vikas Aggarwal[2], Alissa Idelson[2], Sigurd Höger[2], John M. Lupton[1], Jan Vogelsang[\*,1]*

[1]Institut für Experimentelle und Angewandte Physik, Universität Regensburg, 93053 Regensburg, Germany

[2]Kekulé-Institut für Organische Chemie und Biochemie, Universität Bonn, 53121 Bonn, Germany

*Corresponding author. E-Mail: jan.vogelsang@physik.uni-regensburg.de

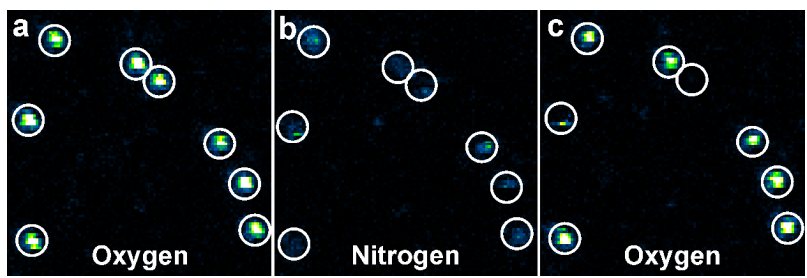

**Figure S1.** Confocal scanning microscopy images of single macrocycle molecules under different conditions. (a) 5 × 5 µm² confocal scan image of single molecules embedded in a 50 nm thick PMMA film under pure oxygen atmosphere. (b) Same area scanned as in (a) but under nitrogen atmosphere. The white circles demonstrate that the same molecules are observed. (c) Same area scanned as in (a) and (b) under subsequent oxygen atmosphere. The molecules are significantly brighter under an oxygen atmosphere due to efficient triplet quenching.

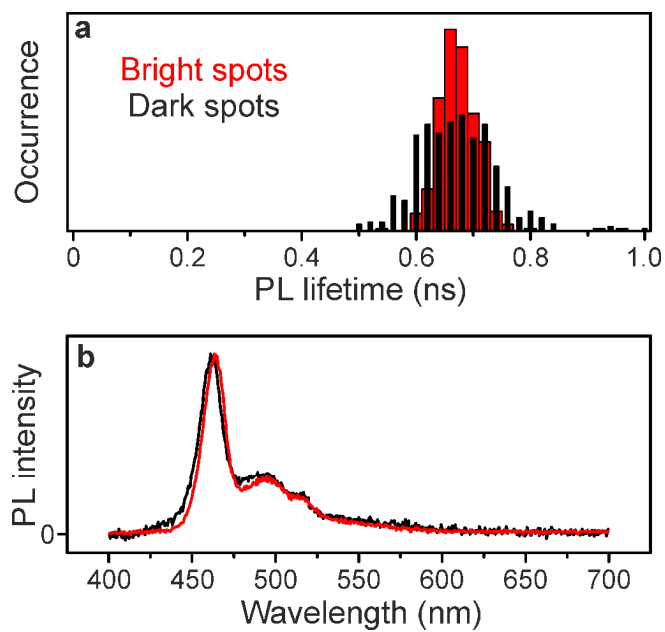

**Figure S2.** Comparison of the photophysics of the bright and dark subpopulation of single molecules embedded in a 250 nm PMMA film. (a) PL lifetime histogram of the bright spots (red bars) and dark spots (black bars). The PL lifetime histogram of the dark spots is slightly broader due to the decreased signal to background ratio. (b) Averaged and normalized single-molecule spectra from 10 bright spots (red curve) and 70 dark spots (black curve). The slight shift of the peak emission from 461 nm (black curve) to 463 nm (red curve) may be attributed to the different environment of molecules embedded deeply in the film and molecules at the film surface.

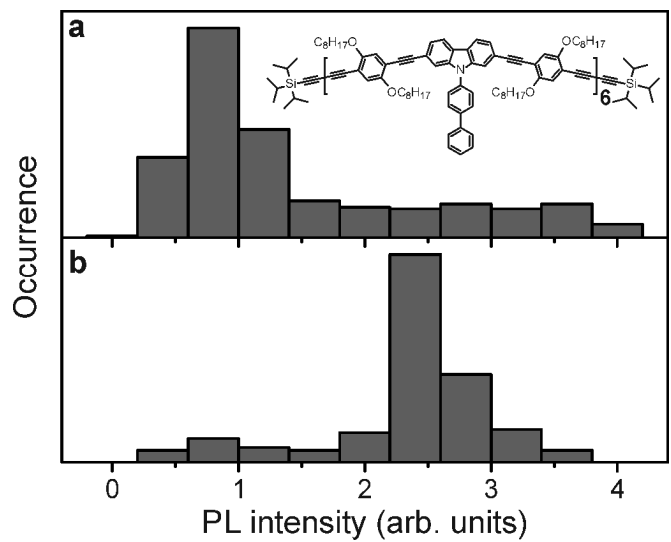

**Figure S3.** PL intensity histograms of linear hexamer oligomers (structure shown in panel (a)) embedded in PMMA obtained by confocal scanning microscopy and subsequent analysis of the diffraction-limited point spread functions. (a) 361 hexamer molecules embedded in a 250 nm thick PMMA film without subsequent SVA. (b) Histogram of 147 hexamer molecules embedded in a 250 nm PMMA film with subsequent SVA. The average number of diffraction limited spots per area is similar for both samples.

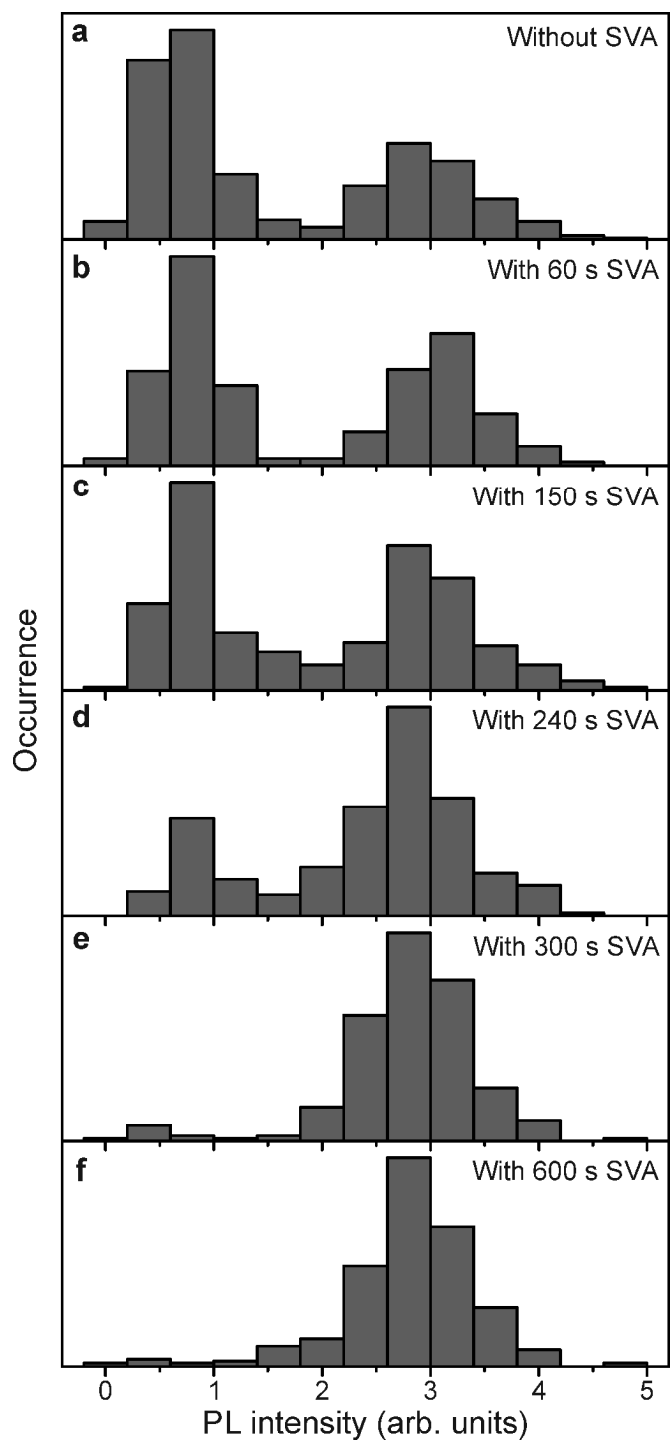

**Figure S4.** PL intensity histograms of single wheel molecules embedded in PMMA obtained by confocal scanning microscopy and subsequent analysis of the diffraction-limited point spread functions. (a) Molecules embedded in a 250 nm thick PMMA film without subsequent SVA yield

a bimodal PL intensity distribution. (b-f) PL intensity histograms of single wheel molecules embedded in 250 nm thick PMMA films subjected to different durations of SVA as denoted in the panels. The average number of diffraction limited spots per area is similar for all samples: 1193 (a), 417 (b), 252 (c), 244 (d), 239 (e) and 324 (f).